\documentstyle[preprint,prl,aps]{revtex}
\begin{document}

\draft

\title{\bf The Quasi-1D S=1/2 Antiferromagnet Cs$_2$CuCl$_4$ in 
a Magnetic Field }

\author{R. Coldea$^1$, D.\ A. Tennant$^1$\cite{tennant_address}, 
R.\ A. Cowley$^1$, D.\ F. McMorrow$^2$, B. Dorner$^3$, Z. Tylczynski$^4$}

\address{$^1$Oxford Physics, Clarendon Laboratory, Parks Road, 
Oxford OX1 3PU, UK\\
$^2$Dept. of Solid State Physics, Ris\o\ National Laboratory,
DK-4000 Roskilde, Denmark\\
$^3$Institut Laue-Langevin, BP 156, F-38042 Grenoble C\'{e}dex 9, France\\
$^4$Institute of Physics, Umultowska 85, 61-614 Poznan, Poland} 

\date{\today}

\maketitle

\begin{abstract}
Magnetic excitations of the quasi-1D $S=1/2$ Heisenberg antiferromagnet
(HAF) Cs$_2$CuCl$_4$ have been measured as a function of magnetic field
using neutron scattering.  For $T<0.62$ K and $B$=0 T the weak inter-chain
coupling produces 3D incommensurate ordering. Fields greater than $B_C$
=1.66 T, but less than the field ( $\simeq$ 8 T) required to fully
align the spins, are observed to decouple the chains, and the system
enters a disordered intermediate-field phase (IFP).  The IFP excitations
are in agreement with the predictions of M\"{u}ller {\it et al.} for the
1D $S=1/2$ HAF, and Talstra and Haldane for the related $1/r^2$ chain (the
Haldane-Shastry model).  This behaviour is inconsistent with linear
spin-wave theory.
\end{abstract}


In recent years there has been a general realization that widely used
semi-classical and mean-field theories of magnetism, such as linear
spin-wave theory, are often inadequate in describing the properties of
low-dimensional quantum magnets \cite{affleck1989}. Research on 1D systems
has benefited from a strong interaction between theory and experiment,
with novel concepts such as fractional excitations and topological energy
gaps being introduced and developed. New magnetic materials continue to be
found and the body of experimental and theoretical work is increasing. In
this Letter we describe a further contribution to this progress by
presenting experiments which show the novel features which occur when a
magnetic field is applied to a quasi-1D antiferromagnet.
        
The canonical 1D quantum mechanical system is the $S=1/2$ Heisenberg
antiferromagnetic chain (HAFC). The low-temperature ground state is a
non-magnetic spin singlet \cite{bethe1931}, and the excitations are
spin-$\frac{1}{2}$ particles known as spinons, as shown by the solution of
the $1/r^2$ Haldane-Shastry model (HSM) \cite{haldane1993}. In a
neutron-scattering process the spin of the system changes by 1 or 0, and
so the neutron-scattering measures a spinon pair continuum. Observations
of this continuum for quasi-1D HAFC's have been made and confirm the
spinon picture \cite{tennant1995I}. The application of a magnetic field
aligns a number of the spins along the field by creating an approximately
regular array of condensed spinons. The triplet excitation continuum in
zero field is then split into separate continua
\cite{muller1981,talstra1994} as illustrated in Fig.\ 1. The positions of
the continua alter as the applied field increases until at high fields $B
\ge B_0$ the spins are aligned parallel along the field and the
excitations have a well-defined magnon dispersion \cite{muller1981}.

The $S=1/2$ HAFC has also been treated by mapping the spin chain into a
system of interacting fermions \cite{pytte1974}. In zero field the
fermion band is half full and the excitation energies can be obtained
using the Hartree-Fock approximation. The transverse excitations are
well-defined and have dispersion relations similar to the lower boundaries
(heavy lines) of Fig.\ 1(a). The application of a field changes the
chemical potential and the Fermi wavevector, such that the transverse 
excitations have incommensurate dispersion relations similar to the 
heavy lines in Fig.\ 1(b).

Although there are a number of excellent realizations of the $S=1/2$ HAFC
available, Cs$_2$CuCl$_4$ is one of the few quasi-one-dimensional systems
with a {\it sufficiently low exchange interaction} that studies of an
applied field {\it throughout} the intermediate-field phase (IFP) can be
made. The IFP is the region of the phase diagram which occurs at low
temperatures ($T \ll J/k_B$), for fields below the ferromagnetic transition
field $B_0$, and when the chains behave independently. For a quasi- as
opposed to pure-1D antiferromagnet this is above a critical field $B_C$, 
the field necessary to decouple the chains, and the IFP region occurs
for $B_C < B < B_0$.

Measurement of magnetic excitations in the IFP is technically challenging
as it requires intense beams of cold neutrons, high magnetic fields, low
temperatures and large single crystals of a quasi-1D material with a low
exchange.  Because of this, previous experiments on quasi-1D $S=1/2$
HAFC's have been exclusively in the province of applied fields which are
low with respect to the exchange, and the difficulty of these experiments
has precluded a comparison with theory \cite{heilmann1978}. In this Letter
we report neutron scattering measurements which, for the first time, span
a large region of the IFP. Although Cs$_2$CuCl$_4$ has a large inter-chain
coupling ($\sim 17 \%$ of the intra-chain value) we argue that the field
decouples the chains and good agreement is found with the theories of
M\"{u}ller {\it et al.} \cite{muller1981}, and Talstra and
Haldane\cite{talstra1994}.

Cs$_2$CuCl$_4$ has an orthorhombic crystal structure \cite{bailleul1991}.
The $S=1/2$ Cu$^{2+}$ spins are coupled into chains running parallel to
the $b$-axis, with four such chains passing through each unit cell.  The
magnetic susceptibility \cite{carlin1985} is consistent with a quasi-1D
$S=1/2$ HAFC with an interaction $J=0.34\pm 0.02$ meV, where the exchange
Hamiltonian for the 1D chains is ${\cal H}_{1D}=J\sum\limits_i$ ${\bf
S}_i\cdot {\bf S}_{i+1}$. The chains are coupled in the $c$-direction by a
small exchange $J^{\prime}=0.175 J$ as depicted in Fig.\ 2(a) (all other
inter-chain exchanges are negligible) \cite{coldea1996}. Below $T_N$=
0.62 K the spins order into a cycloid along the chain direction with
an incommensurate wavevector {\bf q}=(0,0.472,0); the incommensurate
ordering is due to the frustration caused by the staggering of chains with
respect to their neighbours, see Fig.\ 2(a).  A small anisotropy confines
the spins to rotate within a plane containing the $b$-direction and making
a small angle with the $(b,c)$ plane.

The single crystal of Cs$_2$CuCl$_4$ \cite {coldea1996} was mounted with
the $(a,b)$ scattering plane horizontal and a magnetic field was applied
vertically. The sample was cooled using either a dilution refrigerator or
a $^3$He cryostat which provided base temperatures of 0.06 or 0.32 K,
respectively. The inelastic neutron scattering experiments were performed
on the three-axis crystal spectrometer IN14 at Institut Laue-Langevin,
Grenoble, France, while the elastic scattering measurements were made with
the three-axis crystal spectrometer TAS7 at Ris\o\ National Laboratory,
Denmark. Full details of the experiments will be given elsewhere
\cite{coldea_unpublished}.

Strong magnetic reflections were observed at $B$=0 T and $T$=0.32 K in the
($a,b$) plane at positions corresponding to the cycloidal ordering
wavevector ${\bf q}$=(0,0.472,0). When a field was applied along the
crystallographic $c$-axis the intensity of the magnetic Bragg peaks
decreased with increasing $B$ and finally disappeared at the critical
field $B_C$=1.66 T ($T$=0.32 K); for all fields and temperatures the
incommensurate ordering wavevector remained constant. The observed
transition field is much smaller than the field $B_0$ necessary for full
ferromagnetic alignment which was found to be $B_0\simeq$ 8 T from 
susceptibility measurements \cite{coldea_unpublished}.
Fig.\ 2(b) shows a summary of the magnetic phase diagram for
Cs$_2$CuCl$_4$. No antiferromagnetic or incommensurate Bragg peaks were
detected in the intermediate-field phase (IFP), suggesting a decoupling of
the chains.

Fig.\ 2(c) shows the dispersion curves measured at $T$=0.1 K in zero
applied field. The asymmetry in the dispersion relation arises from the
inter-chain coupling and is well accounted for by a calculation of the
spin waves with a Heisenberg exchange $J$ for the interactions along the
chain and $J^{\prime }$ for the coupling between chains
\cite{yoshimori1959}. The solid line in Fig.\ 2(c) represents a fit of the
dispersion relation with $J=0.64\pm 0.02$ meV and $J^{\prime }/J=0.17\pm
0.02$. The latter ratio of exchange interactions is in good agreement with
the value $J^{\prime }/J=0.175$ which gives a stable incommensurate
structure \cite{coldea1996}. The magnitude of $J$ is larger than the value
0.34 meV deduced from the susceptibility measurements \cite{carlin1985} by
using a theory which includes quantum fluctuations. The quantum correction
to the classical spin-wave result for an ideal $S=1/2$ HAF chain is a
multiplication of the energy scale by a factor of $\frac \pi 2$
\cite{dCP1962}, and a value of {$\tilde{J}={\frac{2}{\pi}\ }$0.64=0.41}
meV is similar to the value deduced from susceptibility measurements.
 
The magnetic excitations were studied as a function of field by measuring
the inelastic scattering at low temperatures ($T$=0.06 K).  Fig.\ 3
summarizes the results of constant-$Q$ scans for $Q$=(0,0.75,0). The
zero-field scan shows a well-defined, almost resolution-limited peak
centered at $E$=0.58 $\pm$ 0.01 meV. The solid line is a fit to a Gaussian
peak and the tail at low energies comes from the incoherent scattering.
The intensity of the magnetic excitation decreases on increasing the field
and the lineshape changes above the IFP transition at $B_{\rm C}$=1.66 T.
A good but not unique fit to the lineshape is obtained by two independent
peaks, as shown by the solid line in the figure. The low-energy peak
decreases in energy on increasing the field and has an almost constant
width and intensity for $B_C$ $< B <$ 3 T. In the same field range,
the high-energy peak is at $E$=0.66 $\pm$ 0.02 meV, slightly increased in
energy compared to the peak position in the 3D cycloidal phase. It becomes
broader on increasing the field, but its total integrated intensity is
almost constant.

The inelastic scattering in fields up to 6 T was also measured at
$Q$=(-0.25,2,0), which is equivalent to the zone center for a quasi-1D 
magnetic system with chains along $b^{*}$. The results show an 
inelastic peak for $B >$ 3 T which increases 
in energy and becomes more intense on increasing the
field. This inelastic peak measures the Zeeman energy gap $E_Z=g_c\mu
_BB$, where $g_c$ is the $g$-value along the $c$-axis. A fit to the peak
positions gives $g_c=2.36\pm 0.02$, which is in good agreement with
the value $g_c=2.30\pm 0.01$ determined by previous room temperature EPR
measurements \cite{bailleul1994}.

The results of scanning $Q$ across the zone with 
$E$=0.35 meV are presented in Fig.\ 4. The
zero-field scan has inelastic peaks occurring at positions corresponding
to the intersection of the scan direction with the dispersion relation,
 Fig.\ 2(c). The magnetic inelastic scattering decreases in intensity and
becomes broader on increasing field. The double-peak structure at the zone
boundary $k$=1.5 transforms into a broad feature and the peak on the low-$Q$
side is displaced towards the zone boundary and cannot be distinguished from
the other peak for $B >$ 4 T.

Mean-field ordering and linear spin-wave theory predict that in the case
of a field applied in the plane of spin rotation the IFP is either a
spin-flop (SF) or a cone structure.  Both of these have long-range
antiferromagnetic-like order which was not observed. If the anisotropy
confining the spins in the $(b,c)$ plane is relatively large, the
structure is a SF phase with antiferromagnetic ordering along $b$, and the
transition field cycloidal $\rightarrow$ SF phase is calculated to be
$B_C$=1.425 T \cite{nagamiya1962}. The excitation energy at
$Q$=(0,0.75,0) for isolated 1D chains is field-independent and equal to
0.64 meV \cite{muller1981}.  This is clearly inconsistent with the
observed splitting of the inelastic scattering as a function of field, see
Fig.\ 3. Inter-chain coupling causes some broadening of the scattering,
but cannot account for the results.

In the same mean-field approach, for smaller anisotropies the magnetic
field produces a cone phase in which the spins cant towards the field
direction to gain energy through the Zeeman interaction and the transverse
spin components rotate in the $(a,b)$ plane with the same wavevector {\bf
q} as the zero-field structure in order to minimize the exchange
interactions \cite {nagamiya1962}. The spin-waves in the cone phase have
been calculated \cite {nagamiya1967} and the results for the excitation
energies at $Q$=(0,0.75,0) are presented in Fig.\ 3 by the group of three
vertical thin arrows above the scans. The excitations of the cone
structure also cannot explain the inelastic scattering observed at low
energies with increasing field.

In the case of quantum chains the field produces some spins aligned
parallel to the field direction and approximately regularly distributed
along the chain. If there is no correlation between the positions of the
aligned spins in neighbouring chains, no long-range order will be formed
even in the presence of a small interchain coupling. The IFP thus has
incommensurate-like short-range order and the excitations form a
spinon-magnon continuum according to calculations of Talstra and Haldane
for the $1/r^2$ model, and M\"{u}ller {\it et al.} for the
nearest-neighbor model as shown schematically in Fig.\ 1(b). Measurements
on Cs$_2$CuCl$_4$ for $B>$1.66 T are consistent with the absence of
long-range antiferromagnetic or incommensurate ordering of the IFP phase.
The energies for which the inelastic neutron scattering is expected to be
large for 1D chains with an exchange interaction $\tilde{J}$, according to
calculations of M\"{u}ller {\it et al.} \cite{muller1981}, are shown in
Fig.\ 3 by the thick vertical arrows above the scans. The observed
splitting of the inelastic scattering and its behavior in a field is well
reproduced by the calculation. The strong field dependence of the
low-energy scattering results from the field dependence of the short-range
ordering wavevector as it varies across the zone with increasing field.
The experiments are consistent with the total scattering being constant
for $B_C$ $< B <$ 3 T, in agreement with numerical simulations on
finite quantum chains using the M\"{u}ller Ansatz \cite{muller1981}. The
regions where high intensity is expected in a constant-$E$ scan are
represented by horizontal bars in Fig.\ 4 and the measurements at the
magnetic zone center $Q$=(-0.25,2,0) are also consistent with the calculated
behavior.

In conclusion, we have measured the magnetic ground state and the
excitations in the quasi-1D $S=1/2$ HAF Cs$_2$CuCl$_4$ as a function of
applied field. In the intermediate-field region there is no long-range
antiferromagnetic or incommensurate ordering and the excitations are in
agreement with predictions of M\"{u}ller {\it et al.} for the 1D $S=1/2$
HAF and Talstra and Haldane for the related $1/r^2$ chain.  Mean field
ordering and linear spin-wave theory cannot account for the observed
behavior.  Full details of these and related experiments on Cs$_2$CuCl$_4$
will be reported elsewhere \cite{coldea_unpublished}.

We acknowledge the financial support for the experiments from the EPSRC at
the Institut Laue-Langevin and Oxford, and the EU TMR programme at Ris\o\
National Laboratory. The technical support offered by Frederic Thomas,
Jean-Louis Ragazzoni, Serge Pugol, Alan Brochier and Moritz Lund is 
gratefully
acknowledged. We would like to thank Dr. Kim Lefmann for useful discussions,
and extend thanks to Drs. U.\ Wildgruber and J.\ Axe for
help in lining up the crystal at BNL. R.\ C. is very grateful to the
University of Oxford for the award of a Scatcherd scholarship.

\begin{figure}
\caption{ Spectrum of longitudinal excitations in the 1D S=1/2 HAFC in (a)
zero and (b) an applied field $B = J / g \mu_B$ \protect\cite{muller1981}.
The heavy and light lines are the boundaries of the scattering continua
with the heavy lines indicating where strong scattering is expected.  The
dotted vertical and horizontal lines show the directions of constant-$Q$
and constant-$E$ scans in Figs.\ 3 and 4, respectively. The transverse
modes also show similar effects, but have been omitted for the sake of
clarity \protect\cite{muller1981,talstra1994}.}
\end{figure}

\begin{figure}
\caption{
(a) Magnetic chains and exchange paths in Cs$_2$CuCl$_4$.  (b) Magnetic
phase diagram in a field applied along $c$.  The squares represent the
measured boundary of the 3D cycloidal phase. The Intermediate Field and
Ferromagnetic labels strictly refer to the $T=0$ K line, but are
effectively good descriptions of the ground state at very low temperatures
$T \ll J/k_B$.  (c) Dispersion of the magnetic excitations along the chain
direction at $B=0$ T. The solid line is a fit to the linear spin-wave
dispersion.}
\end{figure}

\begin{figure}
\caption{The intensity observed at $Q$=(0,0.75,0) as a function of energy
and field. Two different scales are used and scans are shifted upwards by
100 (0 T)(right axis), 160 (1.66 T) and 80 (2 T)(left axis). Solid lines
are fits as described in the text, and dotted lines represent the
non-magnetic background. The energies at which the intensity is expected
to be large for quantum chains \protect\cite{muller1981} are shown by
thick vertical arrows above the scans, while the thin arrows indicate the
spin-wave energies for a cone structure. The horizontal bar in the
zero-field scan represents the energy resolution (FWHM).}
\end{figure}

\begin{figure}
\caption{The intensity observed for $E$=0.35 meV as a function of wavevector 
and field. Two different scales are used and scans are shifted upwards 
by 110 (0 T)
(right axis), 150 (3 T) and 70 (4 T)(left axis). Solid lines are guides
to the eye and dotted lines represent the non-magnetic background. 
Horizontal bars above the scans show where high intensity is 
predicted for quantum chains \protect\cite{muller1981}.}
\end{figure}

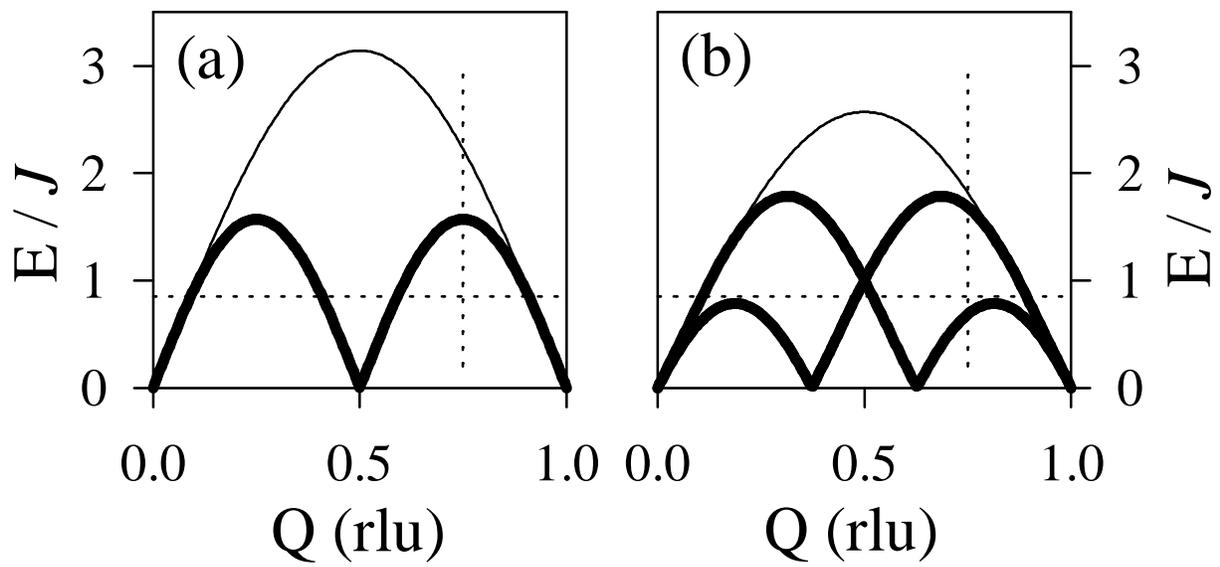
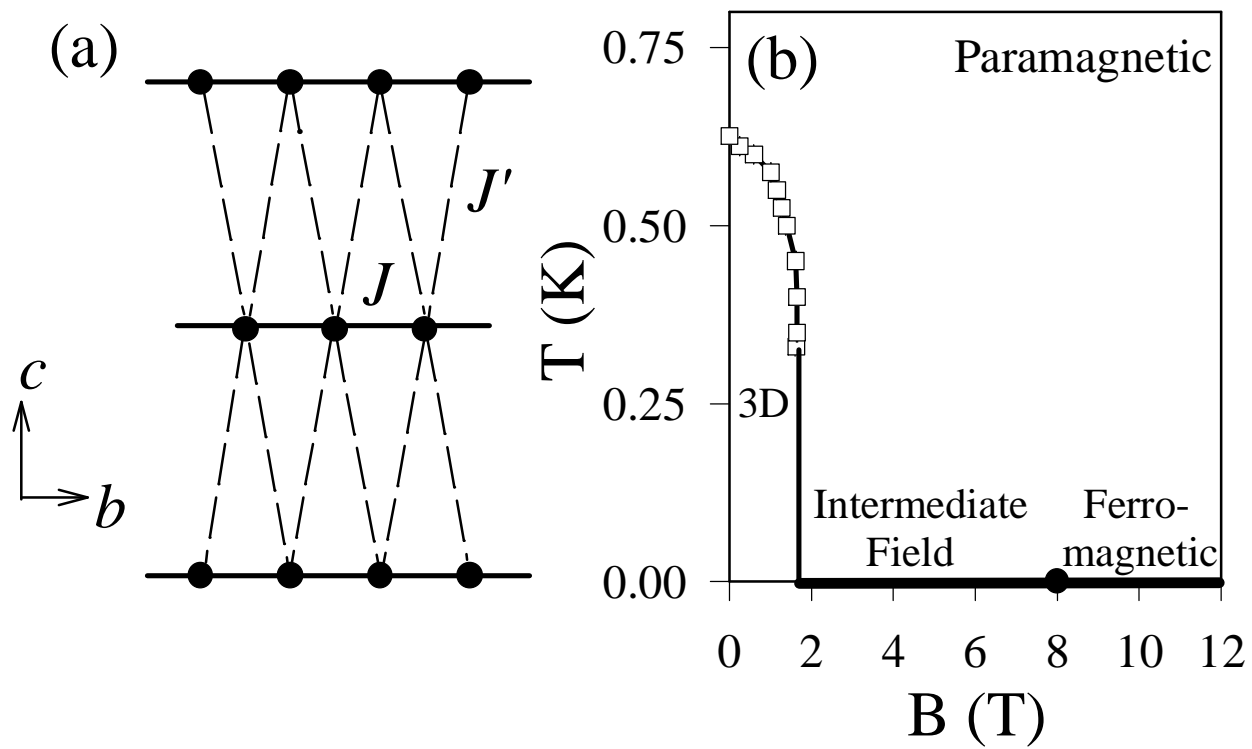
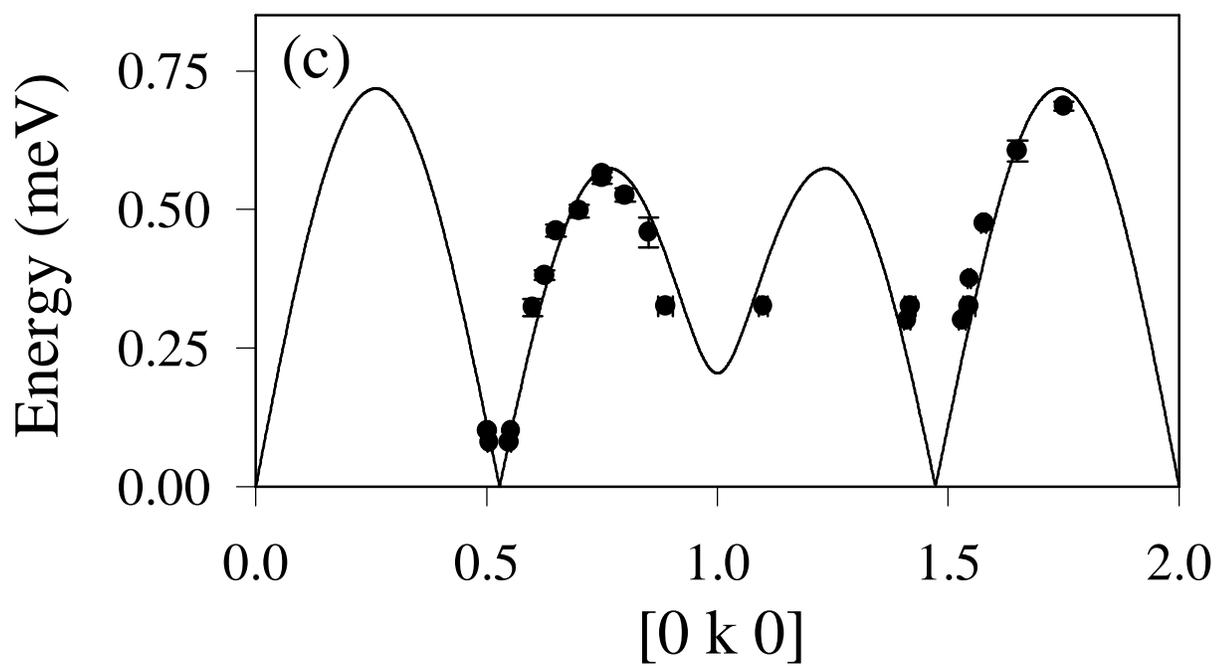
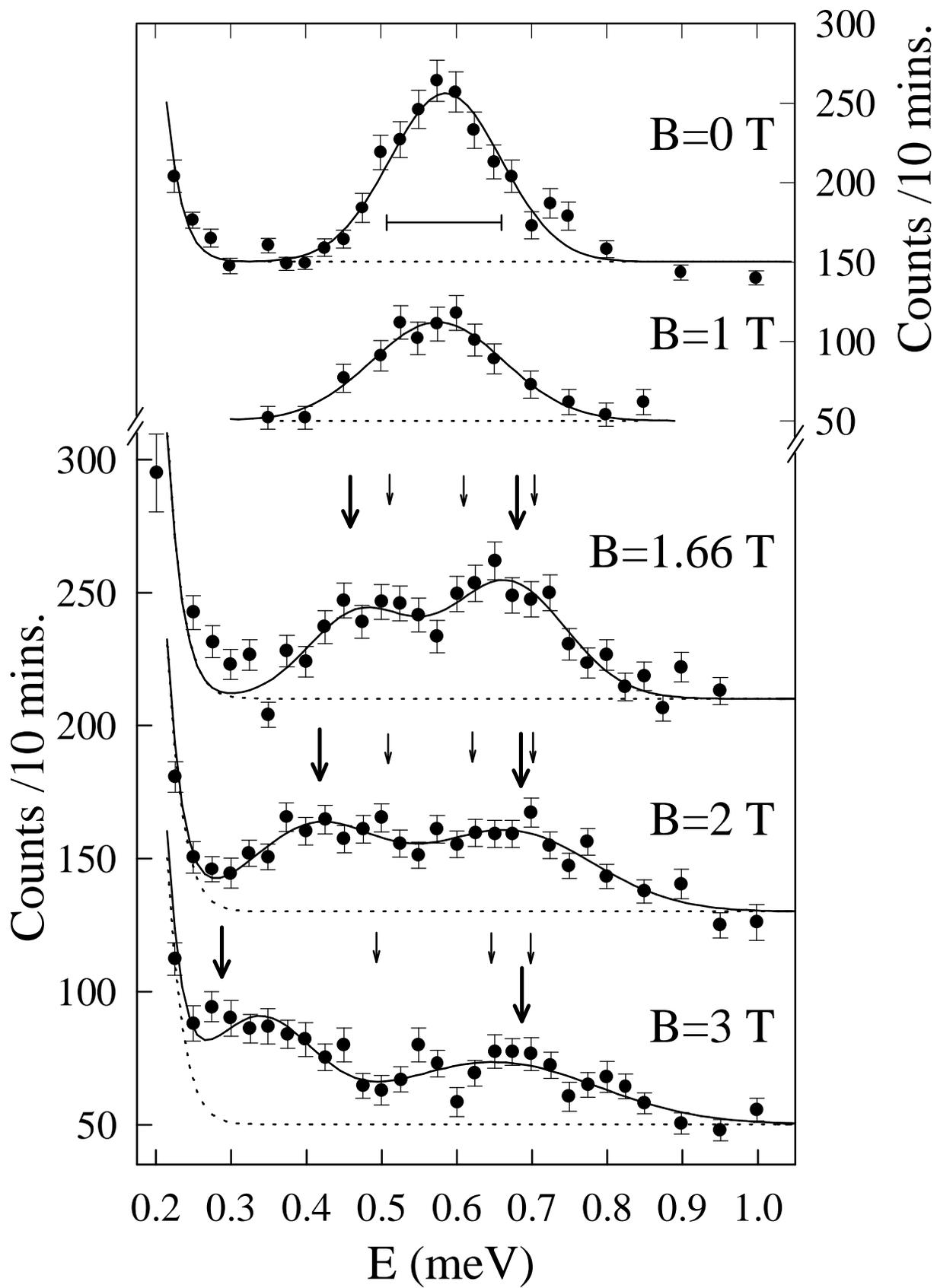
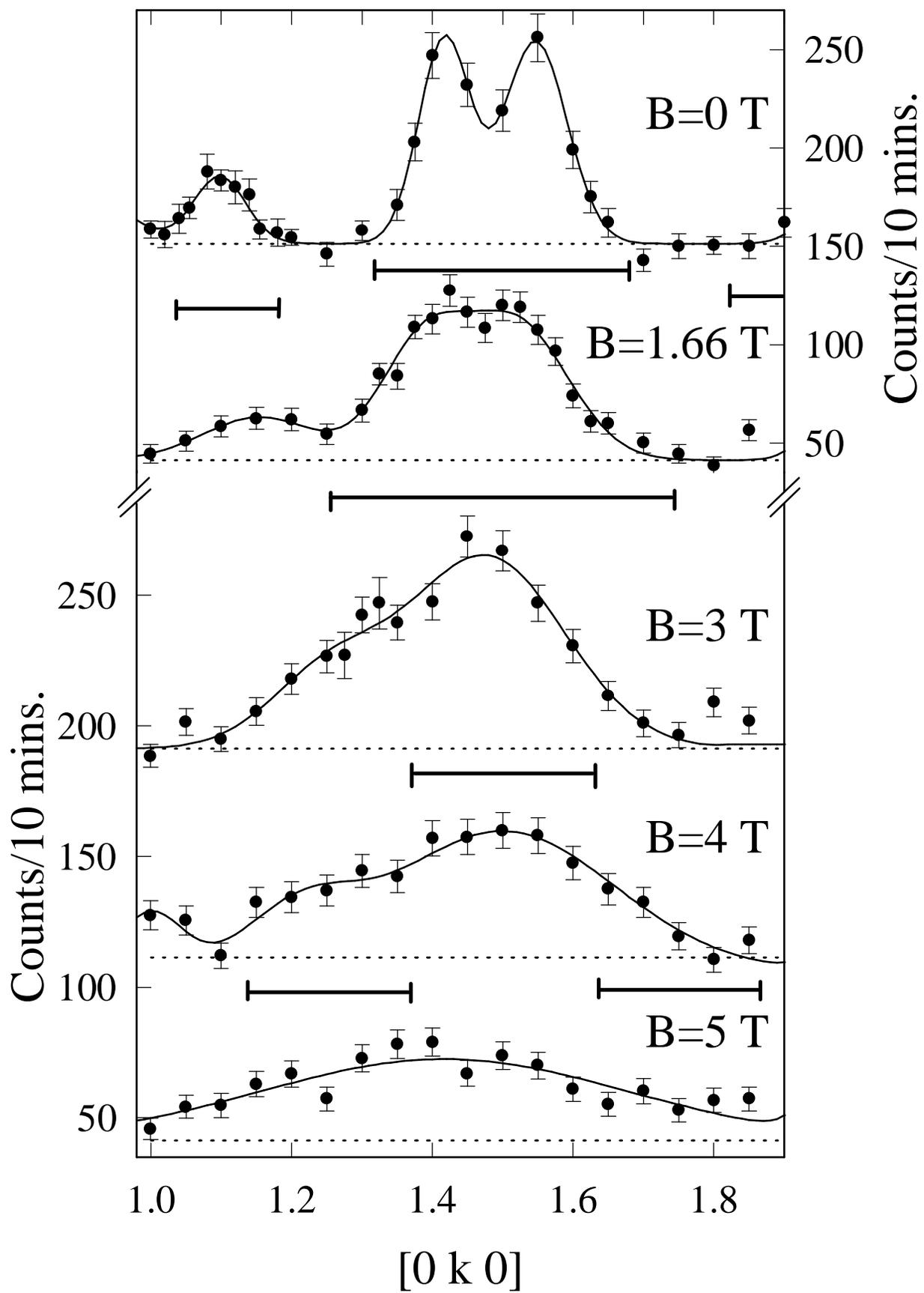
\end{document}